# POSITION PAPER: Credibility of *In Silico* Trial Technologies - A Theoretical Framing

Marco Viceconti, Miguel A. Juárez, Cristina Curreli, Marzio Pennisi, Giulia Russo, and Francesco Pappalardo

***Abstract*— Different research communities have developed various approaches to assess the credibility of predictive models. Each approach usually works well for a specific type of model, and under some epistemic conditions that are normally satisfied within that specific research domain. Some regulatory agencies recently started to consider evidences of safety and efficacy on new medical products obtained using computer modelling and simulation (which is referred to as *In Silico* Trials); this has raised the attention in the computational medicine research community on the regulatory science aspects of this emerging discipline. But this poses a foundational problem: in the domain of biomedical research the use of computer modelling is relatively recent, without a widely accepted epistemic framing for model credibility. Also, because of the inherent complexity of living organisms, biomedical modellers tend to use a variety of modelling methods, sometimes mixing them in the solution of a single problem. In such context merely adopting credibility approaches developed within other research communities might not be appropriate. In this paper we propose a theoretical framing for assessing the credibility of a predictive models for *In Silico* Trials, which accounts for the epistemic specificity of this research field and is general enough to be used for different type of models.**

## I. Introduction

BEFORE a new medical product can be sold in a country, evidence must be provided to the regulatory agency of that country supporting the claim that such new product, if used as expected and under properly controlled conditions, is *safe* (when it does not worsen the health of the recipient) and *effective* (when the product does improve the recipient's health). Historically, evidence of safety and efficacy is provided through controlled experiments. Those experiments involving human volunteers are referred to as *clinical trials*; by contrast those with no humans involved are called *pre-clinical trials*. Some pre-clinical trials involve animals, whereas others are based on cell or tissue cultures, tissues and organs from cadavers, or machineries (bench tests) designed to reproduce the conditions under which the medical product is expected to operate; these are referred to as is *in vitro* tests. So, until recently safety and efficacy were estimated only with controlled experiments *in vitro*, *in vivo* in animals, or *in vivo* in humans. As described in detail in [1], [2], both the USA and European Regulatory Agencies have recently opened in principle to the possibility that some of these regulatory evidences are provided using computer modelling and simulation, what is normally referred to as "*in silico* trials". While for *in vitro* and *in vivo* methods there is an extensive knowledge and a well-established praxis on how to *qualify* them (i.e. how to assess their *credibility* [3] as predictors of the safety and or the efficacy of a new medical product), it is still debated how to assess the credibility of *in silico* methods in a qualification process [4]–[10]. In most cases, methods to assess credibility are merely copied from other research domains, and even applied from time to time to different types of models from those originally developed for. While the first technical standards specifically targeting biomedical applications are appearing [11], there is a clear need for a general theoretical framing on the problem, that can support these efforts.

The aim of this position paper is to propose such theoretical framing for the problem of assessing the credibility of a predictive models for In Silico Trials (ISTs), accounting for the epistemic specificity of this research field and is general enough to be used for different types of models.

## II. Current practices for credibility assessment

### *A taxonomy of predictive models*

As a first step it is useful to categorise the various models used in biomedicine. For the purpose of this position paper, we will categorise them as a function of their knowledge content. In general, a predictive model can be developed by analysing how the quantities to be predicted vary as a function of a set of inputs over a large set of experimental observations (data-driven of **phenomenological models**), or by leveraging some pre-existing knowledge about the physics, chemistry, physiology and biology of the phenomenon being modelled

CORRESPONDING AUTHOR:
Marzio Pennisi
Department of Mathematics & Computer Science
University of Catania, Viale A. Doria, 6
95125 Catania ITALY
Email: mpennisi@dmi.unict.it

M.V. is with the Department of Industrial Engineering, Alma Mater Studiorum - University of Bologna (IT) and the Laboratorio di Tecnologia Medica, IRCCS Istituto Ortopedico Rizzoli, Bologna (IT)

M.J. is with the School of Mathematics & Statistics and Insigneo and Institute for in silico Medicine, University of Sheffield (UK)

C.C. is with Department of Industrial Engineering, Alma Mater Studiorum - University of Bologna (IT)

G.R. and F.P. are with the Dipartimento di Scienze del Farmaco, University of Catania (IT)

(**mechanistic models**). While purely phenomenological models exist, no model is purely mechanistic. Also, there are some modelling approaches that combine mechanistic knowledge and phenomenological evidence (sometime referred to as grey-box models). So, there is a continuum from phenomenological to mechanistic modelling, which is well represented by the degree of mechanistic knowledge used in building each model.

A second important taxonomy is whether the phenomenon is modelled as a continuous or as a series of discrete events. We are not referring here to the need to discretise space and time for obtaining a numerical approximation, but to models that are built assuming the phenomenon being modelled can be described by a finite set of discrete states, whereas a continuous model describes all quantities as continuous in space-time.

*Model Verification & Validation*

Before we go any further, it is important to stress the difference between verification and validation, as the two terms are often confused, even if are they are tailored to different questions.

Simply speaking, verification tries to answer to the question "are we building the system right?", while validation refers to the question "Are we building the right system?". The ASME V&V 40 defines verification as "the process of determining that a computational model accurately represents the underlying mathematical model and its solution from the perspective of the intended uses of modeling and simulation". In other words, we need to test that all the implemented code and the solver approximations of the computational model lead to numerical results that are "sufficiently near" (i.e., taking into account numerical approximations and discretization errors) to the exact analytical solutions of the mathematical formulation of the model at hand.

The goal of Validation is instead to assess how well the computational model represents the reality it is supposed to represent. That is, with validation we check that the model well reproduces the biological, physical, mechanical features of the real phenomenon. To this end, results coming from simulations are compared against results and measurements coming from *in vivo/in vitro* benchmark experiments, usually executed under controlled conditions.

*VV&UQ of engineering models*

In engineering sciences, there are well established practices for assessing model credibility for different industrial sectors, such as aviation, civil, nuclear, etc. These are usually referred to as *Verification*, *Validation*, and *Uncertainty Quantification* (VV&UQ, or sometimes referred to more simply as V&V). A fundamental element in model credibility is what we will call here *Applicability*, which is how far from the conditions under which the model has been validated, the predictions remain credible. In the following we provide a brief description; more details can be found in [12], [13].

Any predictive model can be formalised as:

$$\hat{O} = f(I),$$

where $\hat{O}$ is an estimate of $O$, the quantity we want to predict; $f()$ represents the predictive model; and $I$ the set of input values (or parameters) of the model. In general, $I$ and $\hat{O}$ are scalar, vector, or tensor quantities.

*1) Context of use*

The first step in the credibility assessment for a model is the definition of **the Context of Use** (CoU). This is a detailed explanation of how we plan to use the model in the regulatory process. In particular, we need to define precisely which decisions we would want to make using the model, how the predicted quantity $O$ informs such decisions, what is the structure of $f()$, which inputs $I$ it requires and how they are determined, what is the uncertainty $\sigma(I)$ affecting such inputs, and what is the range of values that $I$ can possibly assume when the model is used in that CoU.

*2) Verification*

The second step is **verification**, which is usually split into *code verification* and *model verification*. Code verification is essentially a quality assurance practice for the software we use to solve the model, and a stability check for the numerical algorithms it implements. Software quality assurance is done with methods typical of software engineering, such as regression tests, or unit tests. Numerical solvers implementations are tested to check their stability, their rates of convergence, their computational efficiency, etc. Commercial codes tend to have their own extensive code verification.

*Model verification*, also called calculation verification or solution verification, mainly aims at evaluating the approximation errors, i.e. the quantification of the error caused by the approximated solution of the mathematical model, $\nu_{f(I)}$. Once the code verification is completed, one can assume that any approximation error is only due to the numerical solution, and not to any coding errors. Methods for the verification of the model depend on the model's structure, and the choice of numerical methods used to solve it. For example, for continuum models these may include simple time-step or grid convergence analysis (e.g., mesh convergence analysis for finite element models).

Normally, for mechanistic approaches, if a particular set of inputs, $I^*$, exists which simplifies $f()$ to the point where an analytical solution is possible, this particular solution $f(I^*)$ is called a benchmark problem, and $\nu_{f(I)}$ can be easily quantified against it. When it is difficult or impossible to formulate a benchmark problem, alternative methods are available such as the *method of the exact solution* and *method of the manufactured solutions* [14].

Another important element in the model verification process is the *parameter exploration*. The model is executed multiple times, each run with a different set of inputs, in order to sample the input space for the range of admissible values, consistent with the problem being modelled. The goals are to: (*i*) ensure that the model runs correctly for all admissible inputs, and (*ii*) that the output vary smoothly with the input. If any of these conditions are not met, it may indicate a problem with the model, for example numerical instabilities.

It should be stressed, that the separation between code verification and model verification can be sometimes a bit artificial. For example, advanced modelling software platforms such as ANSYS [15] provide the modellers with a full-blown programming language to describe the model. However, in complex cases, such languages can be used to describe specific algorithmic behaviours. In such case, the code verification ANSYS provide would not be sufficient, because the additional code written by the user would need itself a code verification.

For completeness, some technical standards, such as V&V 40 [11], consider also as part of the verification, the evaluation of the *use error*, which is due to human errors of the practitioner (for example transcription errors).

*3) Validation*

Once the verification is complete, a validation study must be conducted. If it is possible to carry out an experiment so well controlled that the measured uncertainty of the outcome variable, $\sigma(O)$, is negligible, then the prediction error $\left(|O-\hat{O}|\right)$ can be entirely be ascribed to the model. When this condition is not met, the definition of predictive accuracy becomes more complex, and should be framed within the context of uncertainty quantification and sensitivity analysis.

*4) Uncertainty quantification & Sensitivity Analysis*

Uncertainty quantification (UQ) & Sensitivity Analysis (SA) are essential parts of the validation process. Uncertainty quantification refers to the estimation on how the stochastic error that affects the input propagates through the model into the output:

$$\sigma(\hat{O}) = f(\sigma(I))$$

where $\sigma()$ represents the variability of a quantity over repeated measurements/predictions. Sensitivity analysis is a post-hoc analysis done on the results of the uncertainty quantification, to evaluate which elements of the input set, *I,* are the main drivers of output variability. For a systematic review of the methods used in general engineering, see [16].

*5) Applicability*

The term "Applicability" refers to whether the validation evidences provided are relevant within the CoU of the model. For example, if a metal structure may be occasionally subjected to loads large enough to induce plasticity in the metal, a validation conducted only with low loads where the metal behaves in purely elastic fashion would not provide enough evidence that the model is credible when higher loads are involved, and plasticity may occur. For phenomenological models, such as machine learning, applicability is framed in term of the generalisation error [17]. For mechanistic models it is related to the concept of "limit of Validity" of the theory used to develop the model [18].

*6) Technical standards*

Probably the most important effort in this field is the one of the American Society of Mechanical Engineering (ASME) standardisation committee "V&V: verification and validation in computational modeling and simulation" [19]. It is articulated in a number of sub-committees, each focusing on a specific industrial sector; the one of interest here is the "V&V 40 Verification and Validation in Computational Modeling of Medical Devices" [11].

## III. THE THEORETICAL FRAMING OF CREDIBILITY ASSESSMENT

*From practice to theory*

In the following section, we propose a formal framing for carrying out credibility analyses for *in silico* models in biomedical sciences and highlight the assumptions on which such practices are built upon. Thus, providing a theoretical framework which can be used to evaluate if and when such practices are valid.

Assume we can measure $O_i$, the quantity we want to predict, under certain conditions, $i = 1, \ldots, n$. The predictive error of model $f()$, $A_f$, can be quantified as the root mean square difference between the model prediction and the experimental measurement,

$$A_f = \sqrt{\frac{\sum_{i=1}^{n}(f(I_i) - O_i)^2}{n}}, I_i \in \mathfrak{I},$$

with $\mathfrak{I}$ the set of all admissible inputs defined by all possible conditions that may occur in reality for the phenomenon being modelled.

Because these experiments are usually complex, expensive, and time-consuming, it is almost always true that the number *n* of conditions we can test represents only a small sample of $\mathfrak{I}$. Thus, simply reporting the value of $A_f$ for a small number, *n,* of experiments is not enough to establish the model credibility. We need also to make some assumptions on the distribution and smoothness of $A_f(I)$ over the entire domain $\mathfrak{I}$. But to do this, we need to separate the various sources of error. The predictive error can be decomposed as:

$$O - \hat{O} = \alpha_m = \alpha_I + \varepsilon_f + \nu_{f,I},$$

where $\alpha_m$ is the overall predictive error, which can be decomposed into an **aleatory error, $\boldsymbol{\alpha_I}$**, arising from the intrinsic randomness or variability of the data used to inform the model, and an **epistemic error, $\boldsymbol{\varepsilon_f}$**, that influences both the prior knowledge and the phenomenological derivation. Last, when $f()$ is too complex to be solved analytically, the model prediction is affected also by a numerical **approximation error, $\boldsymbol{\nu_{f,I}}$**, which depends on the structure of the model as instantiated for the inputs $f(I)$. In general, we can only quantify $\alpha_m$ through controlled experiments; to separate the three components of this error, we need to make some assumptions, which when valid enable the VV&UQ process.

The first assumption is that $\alpha_f$ has mean nought. If this is true, then we can write:

$$ave(|O - \hat{O}|) \approx \varepsilon_f + \nu_{f,I}.$$

The second assumption is that the numerical approximation error is negligible, compared to the epistemic error ($\nu_{f,I} = o(\varepsilon_f)$). If this is true, then:

$$ave(|O - \hat{O}|) \approx \varepsilon_f.$$

The third and last assumption is that all the uncertainty affecting $\alpha_m$ is due to the aleatoric component of the prediction error. If this is true, then:

$$\sigma(|O - \hat{O}|) \approx \alpha_I,$$

where $\sigma$ is the variability of the prediction error. Only if these assumptions are justified, we can evaluate the credibility of a predictive model through the standard VV&UQ process:

- First, we need to define the CoU of the predictive model, and the maximum acceptable predictive error.

- Then we need to check that the inputs we provide to our predictive models are accurately measured, in the sense that the measurements we use to quantify them are affected by small systematic errors, if any; this would confirm the first assumption.
- Obtain some values for the output $O$, under conditions for which the epistemic error is known to be nought. The predictive error for such conditions provides an estimate of the numerical approximation error; if this is much smaller than the maximum acceptable predictive error, it would confirm the second assumption.
- With the first two assumptions checked, we can run multiple validation experiments and obtain an estimate of the epistemic error. By looking at how the epistemic error varies for different values of the inputs, we can also inform the applicability of the model, for example defining the limits of validity of the model in terms of input values.
- Last, we need to confirm the third assumption, that all uncertainty comes from the aleatoric component. This is not easy, and it requires some reflections on the internal structure of the model. If the third assumption is confirmed, then we can run an uncertainty quantification study, observing how the variability of the inputs propagates into the model predictions.

A final consideration on the concept of Applicability is needed. The goal is to estimate the predictive error of the model as a function of the input, over the admissible range in the input space, defined by the CoU. Models are usually validated over a limited set of possible input values, which might provide a sparse sampling of such error function. To deem such sparse sampling sufficient for establishing the credibility of the model, we make two additional assumptions.

The first assumption regards the dependence of the model prediction error on the input; the variation of the prediction error should be smooth, i.e. for similar input values the model should produce similar prediction errors, and the difference in the error should be negligible compared to the prediction error of the model for any input; formally, if $I_1 \approx I_2$ in some sense, then:

$$\alpha_M(I_1) - \alpha_M(I_2) \propto o[\alpha_M(I_i)].$$

The second assumption is that the credibility of the model is a decreasing function of the distance between the input values for which the validation was conducted, and the input values for which the model is being used in that specific CoU. As we will discuss in the following section, these assumptions are related to the type of model being used.

## IV. CREDIBILITY ASSESSMENT OF BIOMEDICAL MODELS

### *Why biomedical models are different*

The process to assess the credibility of a predictive model described in the previous sections was developed for general engineering applications, where there is usually fairly robust mechanistic knowledge of the process being modelled; thus, most of these models are physics-based, with a major mechanistic component that describes continuum field problems using systems of ordinary differential equations (ODEs) or partial differential equations (PDEs), solved with well-established numerical methods.

The situation in biomedicine is rather different. For many biological processes the mechanistic knowledge available is partial or totally absent. In many cases the observations used to develop this mechanistic knowledge are affected by considerable errors, and sometimes these are not even quantitative. The complexity of most biological processes is hardly reducible; for example in most cases the assumption of scale separation involves considerable errors [20]. This has forced the researchers in this field to use a wide range of modelling methods, including continuum field models, agent-based models, machine learning models, and even combinations of these, like orchestrated multiscale models.

The standard VV&UQ described before is usually valid for continuum field models, lumped-parameter and compartment models, diffusion-reaction models, etc. But for a few types of models used in biomedicine further discussion is needed.

### *Phenomenological models*

The distinction between statistical and machine learning models is debatable [21], so here we prefer to generically refer to phenomenological models. Is VV&UQ useful to establish the credibility of phenomenological models? Code verification applies to any software artefact, and this includes also phenomenological models. Strictly speaking the concept of model verification does not apply, since there is not mathematical model to solve numerically; some authors use the term model verification to mean validation (e.g. [22]).

Also, the concept of validation is quite different: to develop a phenomenological model we need a number of inputs sets for which the true (i.e. experimentally observed) output set is known, which is another definition of validation set. The difference between the predictions of the final model and the true values of the training set is called *training error*. Unfortunately, because the model is developed by induction, the training error is in no way predictive of the so-called *test error*, which is the error the machine learning model will show when tested against a new validation set. In a sense, the evaluation of the test error is the closest thing to validation for machine learning models.

But the biggest issue with machine learning models is Applicability. First, phenomenological models are highly environment-specific, and any changes of the original field environment may result in a significant different outcome. Thus, they can show a considerable degree of stiffness in the prediction error, when not optimally trained against adversarial perturbations [23]. This means that in some cases the predictive errors may not have the smoothness the concept of applicability requires. But there are deeper issues. Mechanistic models are usually built under some hypotheses (idealisations) which are acceptable only within a range of input conditions, called *limit of validity* of the theory. For these models we can expect the predictive error to be fairly constant (and small) while the inputs are within the limits of validity, and then start to increase the further we go in the inputs space beyond these limits. None of these assumptions can be made for a phenomenological model. Even if we found small prediction errors for all inputs of both the training and test set, in principle nothing will guarantee that for an input not included in these two sets the prediction error is much larger.

One specific issue is called *Concept Drift* [24]. This indicates that the statistical properties of the output, which the model is trying to predict, change over time in unforeseen ways. Of course, in principle the risk of Concept Drift also affects mechanistic models; if the model has been designed and validated over a normal population, and then a characteristic of the population changes, e.g. obesity becomes increasingly common, one might observe a progressive loss of predictive accuracy in all types of models. But in a mechanistic model the inputs are carefully selected to include all factors that might significantly affect the output, so in this case a good model would include body weight among the input variables. In machine learning the input set is selected from the available information to maximise the predictive accuracy on the training set, the risk of not including body weight in the training set if not all patients have normal weight, can be much higher.

Recently, the FDA published a report entitled “Proposed Regulatory Framework for Modifications to Artificial Intelligence/Machine Learning (AI/ML)-Based Software as a Medical Device (SaMD) - Discussion Paper and Request for Feedback” [25]. Even if the target is different (SaMD are software artefacts directly involved in the care process of individual patients), some conclusions are relevant here. In short, FDA advocates for these products the use of total product lifecycle (TPLC) regulatory approach. At risk of oversimplifying, this means that a phenomenological model is never truly validated, and its predictive accuracy should periodically re-tested. In the best-case scenario, every time the model is used to make a prediction, we should later on collect the true value for that case and recalculate the average predictive accuracy; if this degrades below an acceptability threshold the use of the model should be suspended.

### *Agent-based models*

Agent-Based Models (ABM), are mechanistic models, where at least some of the inputs are discrete. Most ABM represent agents as finite-state machines, capable of assuming a limited number of discrete states, moving and interacting in a space-time continuum over which a number of fields are also computed.

In some implementations, only the state transitions are handled as discrete rule-based events, whereas the motion of the agents, and changes in the continuum fields, are expressed by coupled continuum models formulated as systems of ordinary or partial differential equations [26]. In some others, space, time, and field quantities are discretised, so that the entire system can be modelled with discrete-rule based events [27]. But this is a choice made for modelling convenience, as only state variables, and the birth-death of the autonomous agents are epistemically discrete; all the other variables remain conceptually continuous.

Generally speaking, the VV&UQ approach is appropriate to establish the credibility of an ABM. The epistemically continuous parts of the model require extensive model verification, whereas the inherently discrete ones require only code verification, as they involve only algebraic calculations; however, because of the possibility of local instabilities, a parameter exploration is recommended. Validation and uncertainty quantifications can be performed normally. Applicability requires instead some additional cautions. The discrete nature of the ABMs does not ensure the smoothness of the predictive error over the input space, although parameter exploration can reassure of the smoothness of the predictions over the inputs space. The state-transition rules are often based on theories informed by biological observations, which have in some cases faced only limited falsification attempts, and in general are incomplete descriptions of the causal relationships. For all these reasons ABMs predictions made far from the validation sets in the input space should be used with limited confidence.

### *Multiscale and multi-physics models*

Because the assumption of scale separation usually involves significant errors for living organisms, it is becoming quite popular for biomedical problems to use multiscale models. Also, sometimes it is convenient to break down the problem into multiple models, each capturing an aspect of the phenomenon being modelled (sometime these are referred to as Multiphysics models). In both cases, we are not dealing with a single model, but with an orchestration of multiple *component models*. In the case of multiscale model these component models are coupled over space-time through homogenisation / particularisation operators that transform quantities across space-time scales. In the case of Multiphysics models there is also sometime the need to interpose in the orchestration between two models a *transformation* model that remaps certain properties across the specific physics theories.

As discussed in [18], model orchestrations pose peculiar problems when used to falsify theories. The same reasons make it very difficult to discuss if the traditional VV&UQ approach to model credibility is appropriate for this class of models. An in-depth discussion is beyond the scope of this position paper: interested readers can explore the variety of approaches reported in the literature [28]–[34]. But the general philosophy that is emerging is that the credibility of multiscale models should be assessed first by carrying out an individual VV&UQ for each single-scale component model, and another for their multiscale orchestration, with validation experiments relevant to the CoU. In certain complex models, homogenisation and particularisation functions are models themselves, and thus they also should have individual VV&UQ assessments.

Also, uncertainty quantification is challenging with models' orchestrations: due to be combinatory nature of the problem brute force approaches such as Monte Carlo method is most of time prohibitive. In these cases special approaches are required such as Bayesian multi-fidelity schemes [35], or Gaussian processes [36] .

*Applicability for non-mechanistic models*

In the most general case, the model $f()$ is built composing *phenomenological* and *mechanistic* knowledge. Mechanistic knowledge derives from scientific theories that have so far resisted falsification attempts so extensively that the scientific community accept them as operationally true; these are sometimes referred to as *first principles*. A good example of such theory is Newton's second law of dynamics. Phenomenological models are entirely and exclusively based on observational data; these can range from simple statistical regressions, to sophisticated machine learning models. The model $f()$ can then be written as:

$$f(I) = f\big(\mu(I), \psi(I,J)\big).$$

Where $\mu(I)$ is the mechanistic part of the model, $\psi(I,J)$ is the phenomenological model and $J$ is a set of observed data (sometimes referred to as *training set*). The error of the model can then be written as:

$$\varepsilon\big(f(I)\big) = \varepsilon\left(\varepsilon_\mu(I), \varepsilon_\psi(I,J)\right).$$

The question of applicability is related to the shape of $\varepsilon(f(I))$: if the error is nearly constant over a wide range of $I$, we can predict far from the input values used in validation with small risk; if the error varies considerably over $I$, then extrapolation is risky. In most cases, there is a reliable understanding of how the observational error $\varepsilon(I)$ (i.e. instrumentation accuracy) varies across the range of observed values. Since the structure of mechanistic models is explicit (white box models), it is in general possible to estimate the shape of its error from that of the inputs: $\varepsilon_\mu(I) \approx \mu(\varepsilon(I))$. It is instead very difficult to estimate the shape of the error affecting the phenomenological part of the model, $\varepsilon_\psi(I,J)$, due to its dependence on both the inputs and the training set, and the uncertainty on the structure of $\psi(I,J)$ (black box model).

This has an important practical implication: for predominantly mechanistic models error propagation studies make it possible to estimate the shape of the prediction error over the input space, and guide on deciding how far from the validation region the model can still accurately predict; for phenomenological models this is not possible, and thus model applicability should be limited to the range of validated input values.

## V. Possible Context of Use for In Silico Trials

The credibility of a predictive model can be evaluated with respect to CoU aimed at reducing, refining, or replacing an experimental study conducted *in vitro*, or *in vivo* either on animals or humans (clinical trial). While each of these scenarios is potentially relevant for different reasons, we will focus here on the use of predictive models to reduce the number of patients involved in and/or the duration of a clinical trial.

*The standard clinical trial*

The effect of an intervention $i$ is tested clinically by forming a cohort, $\Lambda_n$, of $n$ physical patients (each represented by a patient $j$ descriptor set $I_j$), who are selected to be representative of the target patient population by a set of inclusion and exclusion criteria. Each patient in the cohort is then randomly allocated to an intervention $i$, and the effect measured through the endpoint, $e$. An intervention is considered *effective* if the desired effect on $e$ is observed. For example, in the case of a treatment/no treatment design, the average value of $e$ in the treated subset of the cohort is significantly better. Such conclusion is considered credible only if the statistical test used in the evaluation of the difference in endpoints has large power, i.e. if the statistical power $\Omega(\cdot)$ is large enough (e.g. > 80%). As $\Omega$ typically increases with $n$, larger cohorts can provide stronger statistical evidence.

*In silico-augmented clinical trial*

Assume $f(I_j, p, t)$ is a mathematical model that can predict changes in the endpoint $e$ for each patient, $p$, and for either the *naïve* (i.e. no treatment, $t$=0), or the intervention ($t$=1). The clinical accuracy of such model for patient $I$ can be expressed as:

$$\varepsilon(I_j) = |e - \hat{e}| = \left|e - f(I_j, i)\right|,$$

Where $\hat{e}$ is the prediction of the endpoint $e$. Now, the fundamental idea behind *in silico-augmented clinical trials* is

as follows:

*Proposition 1* (in silico-*augmented clinical trials*)*:* If each patient in a cohort is represented by $I_j$, the effect of the intervention is represented by $e$, and there is a mathematical model $f(I_j, i)$ that can predict $e$ with sufficient accuracy $\varepsilon$ given $I$ and the treatment $i$, we can then generate *virtual* patients by simply creating new $I_j$' sets that are possible in the target population, and use the model $f()$ to do an *in silico* clinical trial on these virtual patients, in order to predict their values of $e$.

The cohort $\Xi_m$ of $m$ virtual patients could be used to supplement cohort $\Lambda_n$ in order to decrease its size for the same statistical power.

The key issue is whether the test of a relevant statistical hypothesis on the effect of $i$, using the virtual cohort, $H(\hat{\Theta}; i, \Xi_m)$, where $\Theta$ parameterises the distribution of endopoints $e$, can be considered a reliable replacement of $H(\Theta; i, \Lambda_n)$.

To answer affirmatively to this question, we should to demonstrate:

1) $\forall I_j \in \Xi_m, \exists\ 0 < n < \infty$, such that $I_j \in \Lambda_n$; i.e. that every virtual patient represented by the descriptor set $I_j$ is a *plausible* virtual patient, in that we would observe that descriptor set in physical patient, if the physical cohort was sufficiently large.
2) the clinical accuracy of the predictor $f()$ estimated on the cohort $\Lambda_n$ is *representative* of the accuracy of the predictor on any other (larger) cohort, and such accuracy is clinically *meaningful*, i.e. $\forall J, e(I_j) < \underline{\varepsilon}$.
3) Last, that $\forall\ 0 < \beta < 1, \exists m(\beta), n(\beta) < \infty$, such that $\Omega\left(H(\hat{\Theta}; i, \Xi_j)\right) > \beta\ \forall j > m(\beta)$, and $\Omega(H(\Theta; i, \Lambda_k)) > \beta\ \forall k > n(\beta)$; i.e. for sufficiently large virtual and physical cohorts, the conclusions of the effect of the intervention assessed from virtual and physical cohorts are identical.

Furthermore, while it is not strictly necessary, in some cases:

4) We may also need to prove that for some relevant measure of discrepancy, $d_H$, there are $m, n < \infty$, such that the distance $d_H(\Xi_m, \Lambda_n) < \delta$, for small $\delta > 0$; i.e. the sets of patient descriptors in the virtual and physical cohorts are sufficiently similar when the cohorts are sufficiently large.

*Virtual Patients plausibility*

To demonstrate that $\Xi_m \subset \Lambda_n$ we need to find other cohorts where $I$ (or more likely one element of $I$) have been observed on a much larger cohort. There are two practical difficulties.

In many cases large enough cohorts are available only for healthy subjects; in that case we need to demonstrate that the range of admissible values observed in the normal population does not change in the diseased population: for example, we can take the male-female ratio from general population studies, because no disease changes the gender.

The second problem is that rarely all elements of $I$ have been observed in a large cohort, so we are forced to take the admissible ranges for different elements, from different populations. This is fine, as far as we can demonstrate that the elements of $I$ are independent, i.e. the variation of one element does not affect the variation of the others. Alternatively, we need to estimate the joint distribution for this set, which might not be easy.

*Meaningful clinical accuracy*

Showing $e(I) \leq \underline{\varepsilon}$ is in general not easy. The first problem is how the clinical accuracy is estimated. In the simplest cases, both $I$ and $e$ are observed at the same time, but in some cases $f()$ is a risk predictor, in the sense it predicts an endpoint that can be observed much later using $I$ observed now. A typical example is the predictor of hip fracture risk, where models predict strength from CT images and currently available clinical data, and this strength is used as a predictor of the risk of fracture, which may occur even five or ten years later. In these cases, three possible approaches can be used:

a) The gold standard is a prospective evaluation with a dedicated clinical trial; a cohort of patients is enrolled, all quantities in the cohort descriptor $I$ measured today are used to predict who will fracture and who will not in the next X years; recall all patients after X years and check who did actually fracture and who did not. This can be a very expensive trial. If X is long many will be lost to the follow-up and, if the incidence of fracture is low, a large number of patients should be enrolled to observe enough fractures.

b) A second approach uses long-term observational databases, where both $I$ and $e$ are observed regularly over X years. Any time after that, one may take the $I$ observed at the beginning of the study and use them to predict how many will have endpoint $e$ after X years. The main limitation is that the data were not collected specifically to validate a specific model. Thus, most likely the information available is not exactly what $f()$ needs to be properly informed, so it is not certain how much of the observed inaccuracy is due to the model, and how much of this is due to poor quality of the data.

c) If a cohort can be formed with a portion of the patients having the endpoint over threshold now (i.e. they have a fracture), and the rest not, and we measure the necessary $I$ now, we will be able to check how good $f()$ is in separating the positive and the negative cases; this is sometimes called *stratification accuracy*. A good stratification accuracy is a necessary but not sufficient condition for a good predictive accuracy; if a predictor has a poor stratification accuracy for sure it will also have a poor clinical accuracy in prospective evaluations; but a good stratification accuracy does not automatically ensure an equally good prospective accuracy.

*Representativeness of the accuracy: power and applicability*

If the physical cohort used to evaluate the clinical accuracy is large enough, this problem can be formulated as a statistical power problem. But even when there are enough data to achieve sufficient statistical power, a more general problem of applicability remains. In order to be useful, a model should be able to make predictions for input values that are different from those used to assess its accuracy; but we do not know the predictive accuracy of the model for those new inputs. Considerations on the general regularity of physical quantities, and about the assumption that model accuracy should degrade smoothly in the sense that predictions made for similar inputs should present similar predictive accuracy, allow to assume that a degree of extrapolation is possible, in the sense that the model can be considered reliable even when used to predict for inputs different from those observed in the clinical validation cohort (Applicability).

*Similarity measures in cohort expansion*

The concept of Applicability does have another important implication for the use of predictive models for cohort expansion. A possible approach is described in [37], [38]. The idea is:

1. Let $L(e \in \Lambda)$ be the likelihood of an intervention $i$ given the responses $e$ observed over the physical cohort $\Lambda$. Assuming conditional independence, $L(e \in \Lambda) \propto \prod_{j=1}^{k} P(e|i_j)$; in other words, the likelihood is proportional to the product of the probability distribution of effect $e$, for each intervention level. In the simplest case of a single intervention level, $L(e \in \Lambda) \propto P(e_{treat}) \times P(e_{no\ treat})$.

2. According to Bayes theorem, the posterior probability that the intervention $i$ produces the desired effect $e$ is $P(e|i) = \frac{L(i|e) \times \pi(i)}{P(e)}$, where $\pi(i)$ is the prior probability, and $P(e)$ is the marginal probability of $e$. $L(i|e)$ and $P(e)$ are observed experimentally, and thus their precision is limited by the numerosity of $\Lambda$. But if we can elicit $\pi(i)$ from the virtual patient cohort $\Xi$, whose size $m$ can be freely increased, we can boost the power of $P(e)$.

3. In most cases, the confidence we have on $L(i|e)$ is greater than that we have on $\pi(i)$, which is elicited, but its power is insufficient because the cohort of physical patients $\Lambda$ is too small. Adding a prior may improve the posterior precision, but the risk with this approach is that if $m >> n$, the posterior probability $P(e|i)$ might be influenced excessively by the virtual patients. So, we look for the smallest number of virtual patients sufficient to attain the desired power.

Regarding the selection of the number of virtual patients, Haddad et al. [38] suggest to begin by fixing the maximum number (strength) of virtual patients, $n_{max}$, and adjust the actual number used, $n_0 = n_{max}F$, by multiplying it for a discount factor,

$$F = (p|\kappa, \lambda) = 1 - exp\left[-\left(\frac{p}{\lambda}\right)^{\kappa}\right],$$

which depends on a measure of compatibility of the virtual and real patients (they call it a Bayesian $p$-value) and two parameters, $(\kappa, \lambda)$, controlling the penalisation. These parameters are fixed on an ad hoc basis.

This so-called $p$-value is arguably a measure of (stochastic) dissimilarity (i.e. compares the posterior distributions) between the parameter of interest, $\theta$, from the trial using a minimally informative prior on the augmented cohort ($\theta_F$) and the one obtained using the virtual cohort only ($\theta_0$),

$$p = P[\theta_F \leq \theta_0],$$

estimated using the simulated data.

They argue that as $p \to 1$, the compatibility of the virtual cohort is greater and thus $n_0 \to n_{max}$. This seems to indicate they regard larger values of $\theta$ as indicating compatibility with the real patients. Hence, if the posterior distribution using the real data and a flat prior has more mass towards smaller values of $\theta$, compared to the posterior distribution using the virtual cohort, the method indicates compatibility with the real data and allows for a larger contribution from the simulated patients.

## VI. Conclusions

The aim of this position paper is to propose a theoretical framing for assessing the credibility of a predictive models for In Silico Trials, which accounts for the epistemic specificity of this research field and is general enough to be used for different types of models. We also proposed mathematical formalisms for one particular application of in silico methods, where validated predictive models are used to augment clinical trials on physical cohorts by simulating the clinical trial of a new intervention on virtual cohorts. A recap is proposed in Table 1. The potential scope for *in silico* methods is of course much broader: in principle *in silico* methods can be used also for reducing, refining, or replacing experimentations done *in vitro*, or *in vivo*.

The approach that most practitioners have adopted so far is conservative and aims to frame *in silico* methods within the current structure and logic of testing new medical interventions. In most cases we seek to demonstrate equivalence with the experimental method we aim to replace, but the general logic behind remains the same. The development of *in silico* methods has much deeper implications.

<table>
<tr><th colspan="3">In Silico Augmented Clinical trials: credibility</th></tr>
<tr><td colspan="3"></td></tr>
<tr><td colspan="3">Goal: Create a cohort $\Xi_m$ of m virtual patients to supplement the cohort $\Lambda_n$ of n real patients in order to decrease its size for the same statistical power</td></tr>
<tr><td colspan="3"></td></tr>
<tr><th>Requirement</th><th>Mathematical Formulation</th><th>Proposed methodologies to assess requirement</th></tr>
<tr><td>1) Assess the credibility of virtual patients</td><td>$\forall I_j \in \Xi_m, \exists\, 0 < n < \infty{:}\ I_j \in \Lambda_n;$</td><td>Find other cohorts where I (or more likely one element of I) have been observed on a much larger cohort.<br>Weaknesses:<br>1) large enough cohorts are available only for healthy subjects; demonstrate that the range of admissible values of the normal population does not change in the diseased population<br>2) Rarely all elements of I are observed in a large cohort; take the admissible ranges for different elements from different populations, if the elements of I are independent. Alternatively, try to estimate the joint distribution for the set.</td></tr>
<tr><td>2)Assess that the clinical accuracy is meaningful</td><td>$\forall J, e(I_j) < \underline{\varepsilon}.$</td><td>Possible approaches:<br>1) Use prospective evaluation with a dedicated clinical trial; Weakness: Evaluation can be done only at the end of the trial.<br>2) Use of long-term observational databases. Weakness: Uncertainty about observed inaccuracy (i.e., is due to the model or to inadequacy of data coming from other experiments?).<br>3) Use stratification accuracy. Weakness: Necessary but not sufficient condition for a good predictive accuracy.</td></tr>
<tr><td>3)Assess that, for successfully large cohorts, the conclusions of the effect of the intervention assessed from virtual and physical cohorts are identical</td><td>$\forall\ 0 < \beta < 1, \exists m(\beta), n(\beta) < \infty{:}$<br>$\Omega\left(H(\hat{\Theta}; i, \Xi_j)\right) > \beta\ \forall j > m(\beta),$<br>and<br>$\Omega\left(H(\Theta; i, \Lambda_k)\right) > \beta\ \forall k > n(\beta);$</td><td>Formulate the problem as a statistical power problem, when the cohorts are large enough. Weakness: Not sufficient to make predictions for input values outside those used to assess accuracy; Assumptions about regularity of physical quantities and smoothness of model accuracy degradation may allow application on inputs outside the one used in the clinical validation cohort to some extent (Applicability).</td></tr>
<tr><td>4) (Optional) Assess that sets of patient descriptors in the virtual and physical cohorts are sufficiently similar when the cohorts are sufficiently large.</td><td>$\exists\, m, n < \infty{:}\ d_H(\Xi_m, \Lambda_n) < \delta,$<br>for small $\delta > 0$ (for some $d_H$)</td><td>Proceed as for point 3.</td></tr>
<tr><th colspan="3">Cohort Expansion</th></tr>
<tr><td colspan="3">Methodology: Elicit the prior probability $\pi(i)$ that an intervention i produces a desired effect e from the virtual patient cohort $\Xi_m$, whose size $m$ can be freely increased, to boost the power of $P(e)$, marginal probability of $e$ and to better estimate the posterior probability $P(e|i)$. Weakness: if $m \gg n$, the posterior probability $P(e|i)$ might be influenced excessively by the virtual patients. In this case it is possible use the approach by Haddad et al. to estimate the size of the virtual cohort.</td></tr>
<tr><td>$I_j$ patient j in a given cohort;<br>$\Xi_m$ cohort of virtual patients;<br>$\Lambda_n$ cohort of real patients;<br>$e$ effect of a given intervention (endpoint)<br>$\hat{e}$ is the prediction of the endpoint $e$<br>i intervention (i.e., drug or medical device)<br>$d_H$ a given measure of discrepancy</td><td colspan="2">$\varepsilon$ accuracy of the mathematical model $f(I_j, i)$ in reproducing e<br>$H(\hat{\Theta}; i, \Xi_m)$ statistical hypothesis on the effect i, using the virtual cohort<br>$H(\Theta; i, \Lambda_n)$ statistical hypothesis on the effect i, using the real cohort,<br>$\hat{\Theta}$ the set of endpoints of $\hat{e}$<br>$\Theta$ the set of endpoints of $e$<br>$\Omega\left(H(\Theta; i, \Lambda_n)\right)$ conclusions on the effects of a given intervention i over a given cohort $\Lambda_n$, given the endpoints $\Theta$</td></tr>
</table>

**Table 1.** Summary of the components of credibility in an in silico clinical trial.

Since 1025 (when Ibn Sina completed the Canon of Medicine), the whole basis of the performance assessment of medical interventions has been that the mechanisms that regulate the disease and the effect of an intervention on it are too complex to be explored mechanistically and can be investigated only through observation. The intervention is trailed on a number of closely monitored patients, and if we do not observe any severe adverse effect, and the intervention shows to be effective in these patients, we authorise the use of such intervention on everyone.

In other industrial sectors safety is largely ensured through modelling and simulation. It is inconceivable that a new large passenger airplane is produced in a few copies, which are run on test routes with real passengers, while we monitor if anything goes wrong; no passenger can board a new airplane design before its safety has been thoroughly tested through a combination *in vitro* and *in silico* methods.

This is because for every known failure mode for an airplane a predictive model can be developed. As *in silico* methods develop and mature, both in breath of scope and credibility, there will be a growing opportunity for replacement, maybe computer simulations replacing physical experiments entirely. Probably the low-hanging fruits here are those situations where the animal model, or the standard clinical trial are already known to be poor predictors of the real-world safeness and/or effectiveness. As this revolution occurs, it will become vital to have a rigorous theoretical framing for the processes used to assess the credibility of *in silico* methods. We hope this paper can provide a first contribution in this direction.

## Acknowledgment

The authors would like to thank the many people within whom we discussed of the contents of this manuscript, receiving useful advice. Among the others we would like to name Tina Morrision (FDA), Flora Musuamba-Tshinanu (EMA), Jeffrey Bischoff (Zimmer), Marc Horner (Ansys), Markus Reiterer, Mark Palmer and Tarek Haddad (Medtronic). This study was partially supported by the STriTuVaD project (SC1-PM-16-2017- 777123), the MOBILISE-D project (IMI2-2017-13-7-820820), the PRIMAGE project (SC1-DTH-07-2018- 826494) and by the UK Engineering and Physical Sciences Research Council through the MultiSim Project (EP/K03877X/1).